\documentstyle[12pt,epsf,subfigure]{article}
\newcommand{\cd}{\makebox[0.08cm]{$\cdot$}}

\title{Three-boson relativistic bound states with zero-range interaction}
{\author{J. Carbonell$^a$ and V.A. Karmanov$^b$ \\
{\small \em $^a$Institut des Sciences Nucl\'eaires, 53 avenue des Martyrs,
38026 Grenoble, France}
\\{\small \em  $^b$Lebedev Physical Institute, Leninsky Prospekt 53, 119991
Moscow, Russia}}}
\begin{document}
\maketitle
\bibliographystyle{unsrt}

\begin{abstract}
{For the zero-range interaction providing a given mass $M_{2}$ of the two-body
bound state, the mass $M_3$ of the relativistic three-boson state is
calculated. 
We have found that the three-body system exists only when $M_2$ is greater than
a critical value $M_{c}\approx 1.43\,m$ ($m$ is the constituent mass).  For
$M_2=M_{c}$ the mass $M_3$ turns into zero and for
$M_2<M_{c}$ there is no solution with real value of $M_3$.}
\end{abstract}


\bigskip
Zero-range two-body interaction provides an important limiting case which
qualitatively reflects characteristic properties of nuclear \cite{Nuclear}
and atomic \cite{Atomic}
few-body systems.
In the nonrelativistic three-body system it generates the Thomas collapse
\cite{thomas}. The latter means that the three-body binding energy tends to
$-\infty$, when the interaction radius tends to zero. Several ways to
regularize this interaction have been proposed in the literature
\cite{FRED,FJ}. 

When the binding energy or the exchanged particle mass is not negligible in
comparison to the constituent masses, the nonrelativistic treatment becomes
invalid and must be replaced by a relativistic one. Two-body calculations show
that in the scalar case, relativistic effects are repulsive (see e.g.
\cite{MC_PLB_00}). Relativistic three-body calculations with zero-range
interaction have been performed in a  minimal relativistic model \cite{noyes}
and in the framework of the Light-Front Dynamics \cite{tobias}. In these works
it was concluded that, due to relativistic repulsion, the three-body binding
energy remains finite and the Thomas collapse is consequently avoided.

In the present paper we reconsider, in the Light-Front Dynamics approach,
the problem of three equal mass ($m$) bosons interacting via zero-range forces.
We will show that instead of the Thomas collapse, its relativistic counterpart takes place.
Namely, when the two-body bound state mass $M_2$ decreases, the mass $M_3$ of
the three-body system decreases as well and vanishes at some critical value
of $M_{2}=M_{c}\approx 1.43\,m$.
For $M_{2}<M_{c}$ there are no solutions with real value of $M_3$
what means -- from physical
point of view -- that the three-body system collapses.

Our starting point is the explicitly covariant
formulation of the Light-Front Dynamics (see for a review \cite{cdkm}).
The wave function is defined on the light-front plane given by the equation
$\omega\cd x=0$, where $\omega$ is a four-vector with $\omega^2=0$, determining
the light-front orientation. In the particular case $\omega=(1,0,0,-1)$ we
recover the standard approach \cite{BPP_PR_98,BKT_NPB_79}.

\begin{figure}[!ht]
\begin{center}
\mbox{\epsfxsize=15.cm\epsffile{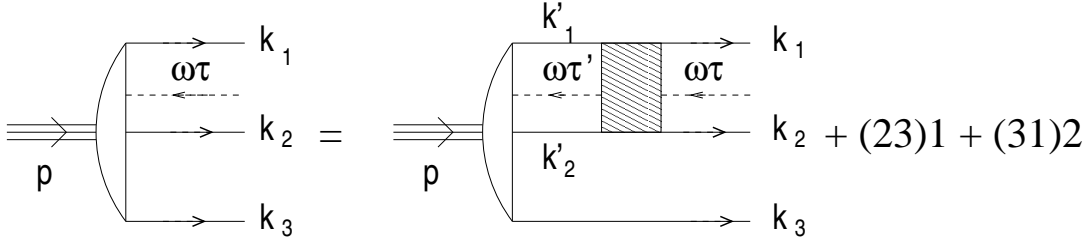}}
\caption{Three-body equation for the vertex function $\Gamma$.\label{fig1}}
\end{center}
\end{figure}

The three-body equation is represented graphically in figure \ref{fig1}.
It concerns the vertex function $\Gamma$,
related to the wave function $\psi$ in the standard way:
$$
\psi(k_1,k_2,k_3,p,\omega\tau)=\frac{\Gamma(k_1,k_2,k_3,p,\omega\tau)}
{{\cal M}^2-M^2_3},\quad
{\cal M}^2=(k_1+k_2+k_3)^2=(p+\omega\tau)^2.
$$

All four-momenta are on the corresponding mass shells ($k_i^2=m^2,
p^2=M_3^2,(\omega\tau)^2=0$) and satisfy the conservation law
$k_1+k_2+k_3=p+\omega\tau$ involving  $\omega\tau$. The four-momenta
$\omega\tau$ and $\omega\tau'$ are drawn in figure \ref{fig1} by dash lines.
The off-energy shell character of the wave function is ensured by the scalar
variable $\tau$.  In the standard approach, the  minus-components of the momenta
are not conserved and the only non-zero component of $\omega$ is
$\omega_-=\omega_0-\omega_z=2$. Variable $2\tau$ is just the non-zero
difference of non-conserved components $2\tau=k_{1-}+k_{2-}+k_{3-}-p_-$.

Applying to figure \ref{fig1} the covariant light-front graph techniques \cite{cdkm}, we find:
\begin{eqnarray}\label{eq3}
&&\Gamma(k_1,k_2,k_3,p,\omega\tau)=
\frac{\lambda}{(2\pi)^3} \int \Gamma(k'_1,k'_2,k_3,p,\omega\tau')
\nonumber\\
&&\times
\delta^{(4)}(k'_1+k'_2-\omega\tau'-k_1-k_2+\omega\tau)
\frac{d\tau'}{\tau'}\,
\frac{d^3k'_1}{2\varepsilon_{k'_1}}\,
\frac{d^3k'_2}{2\varepsilon_{k'_2}}
\;+\; (23)1+(31)2,
\end{eqnarray}
where  $\varepsilon_{k}=\sqrt{m^2+\vec{k}^2}$.
For zero-range forces, the interaction kernel is replaced by a constant $\lambda$.
In (\ref{eq3}) the contribution of interacting pair
12 is explicitly written while
the contributions of the remaining pairs are denoted by $(23)1+(31)2$.

Equation (\ref{eq3}) can be  rewritten in variables $\vec{R}_{i\perp},x_i,$
($i=1,2,3$),
where $\vec{R}_{i\perp}$ is the spatial component of the four-vector
$R_i=k_i-x_ip$ orthogonal to $\vec{\omega}$
and $x_i={\omega\cd k_i\over\omega\cd p}$ \cite{cdkm}.
For this aim we insert in r.h.-side of (\ref{eq3}) the unity integral
$$
1=\int
2(\omega\cd k'_3)\delta^{(4)}(k'_3-k_3-\omega\tau_3)d\tau_3\,
\frac{d^3k'_3}{2\varepsilon_{k'_3}}
$$
and recover the usual three-body space volume
which, expressed in the variables $(\vec{R}_{i\perp},x_i)$, reads
\[
\int
\delta^{(4)}(\sum_{i=1}^3 k'_i-p-\omega\tau')
\prod_{i=1}^3{d^3k'_i\over2\varepsilon_{k'_i}} 2(\omega\cd p) d\tau'
=\int\delta^{(2)}(\sum_{i=1}^3\vec{R'}_{\perp i})
\delta(\sum_{i=1}^3 x'_i-1)2\prod_{i=1}^3{d^2R'_{\perp i}dx'_i\over2x'_i}.\]

The Faddeev amplitudes $\Gamma_{ij}$ are introduced in the standard way:
$$\Gamma(1,2,3)=\Gamma_{12}(1,2,3)+\Gamma_{23}(1,2,3)+\Gamma_{31}(1,2,3), $$
and equation (\ref{eq3}) is equivalent to a system of three coupled equations.
With the symmetry relations $\Gamma_{23}(1,2,3)=\Gamma_{12}(2,3,1)$ and
$\Gamma_{31}(1,2,3)=\Gamma_{12}(3,1,2)$,
the system is reduced to a single equation for one of the amplitudes, say $\Gamma_{12}$.

In general, $\Gamma_{12}$ depends on
all  variables ($\vec{R}_{i\perp},x_i$), constrained by the relations
$\vec{R}_{1\perp}+\vec{R}_{2\perp}+\vec{R}_{3\perp}=0$, $x_1+x_2+x_3=1$,
but for  contact kernel it depends only on $(\vec{R}_{3\perp},x_3)$ \cite{tobias}.
Equation (\ref{eq3}) results into:
\begin{equation}\label{eq24}
\Gamma_{12}(\vec{R}_{\perp},x)
=\frac{\lambda}{(2\pi)^3}\int \left[\Gamma_{12}(\vec{R}_{\perp},x)
+2\Gamma_{12}\left(\vec{R'}_{\perp}-x'\vec{R}_{\perp},\;
x'(1-x)\right)\right]
\,
\frac{1}{s'_{12}-M_{12}^2}
\frac{d^2R'_{\perp}dx'}{2x'(1-x')},
\end{equation}
in which
\[s'_{12}=(k'_1+k'_2)^2= \frac{{R'}^2_{\perp}+m^2} {x'(1-x')}  \]
is the effective on shell mass squared of the two-body subsystem,
whereas $M^2_{12}=(k'_1+k'_2-\omega\tau')^2=(p-k_3)^2$
corresponds to its off-shell mass.
It is expressed through $M_3^2,R_{\perp}^2,x$ as
\begin{equation}\label{eq17}
M^2_{12}=(1-x)M_3^2-\frac{R_{\perp}^2+(1-x)m^2}{x}.
\end{equation}
These on- and off-shell masses $s'_{12}$ and $M^2_{12}$ differ from each
other, since $k'_1+k'_2+k_3\neq p$.
On the energy shell, at $\tau'=0$, the value $M^2_{12}$ turns into
$s'_{12}$, what is never reached for a bound state.

Since the first term $\Gamma_{12}(\vec{R}_{\perp},x)$ in the integrand does not
depend on the integration variables, we can transform (\ref{eq24}) as:
\begin{equation}\label{eq25}
\Gamma_{12}(\vec{R}_{\perp},x)
=\frac{1}{\lambda^{-1}-I(M_{12})}
\frac{2}{(2\pi)^3}\int
\Gamma_{12}\left(\vec{R'}_{\perp}-x'\vec{R}_{\perp},x'(1-x)\right)
\frac{1}{s'_{12}-M_{12}^2}\frac{d^2R'_{\perp}dx'}{2x'(1-x')},
\end{equation}
where
\begin{equation}\label{eq26}
I(M_{12})=\frac{1}{(2\pi)^3}\int \frac{1}{s'_{12}-M_{12}^2}
\frac{d^2R'_{\perp}dx'}{2x'(1-x')}.
\end{equation}
The integral (\ref{eq26}) diverges logarithmically and we
implicitly assume that a cutoff $L$ is introduced.

The value of $\lambda$ is found by solving the two-body problem
with the same zero-range interaction under the
condition that the two-body bound state mass has a fixed value
$M_{2}$. From that we get $\lambda^{-1}= I(M_{2})$ with  $I$ given by
(\ref{eq26}). It also diverges when the
momentum space cutoff $L$ tends to infinity (or, equivalently, the interaction
range tends to zero). However, the difference $\lambda^{-1}-I(M_{12})=
I(M_{2})-I(M_{12})$ converges in the limit $L\to\infty$.
The factor $F(M_{12})=
1/[I(M_{2})-I(M_{12})]$ gives the two-body off-shell scattering amplitude,
depending on the off-shell two-body mass $M_{12}$, without any regularization.  
For $0\leq M_{12}^2<4m^2$ the calculation gives:
$$
F(M_{12})=\frac{8\pi^2}{\frac{\displaystyle{\arctan y_{M_{12}}}}
{\displaystyle{y_{M_{12}}}}
-\frac{\displaystyle{\arctan y_{M_{2}}}}{\displaystyle{y_{M_{2}}}}},
$$
where
$ y_{M_{12}}=\frac{M_{12}}{\sqrt{4m^2-M_{12}^2}}$
and similarly for $y_{M_{2}}$.
If $M_{12}^2<0$, the amplitude obtains the form:
$$
F(M_{12})=\frac{8\pi^2}{\frac{\displaystyle{1}}
{\displaystyle{2y'_{M_{12}}}}\log \frac{\displaystyle{1+y'_{M_{12}}}}
{\displaystyle{1-y'_{M_{12}}}}
-\frac{\displaystyle{\arctan y_{M_{2}}}}
{\displaystyle{y_{M_{2}}}}},
$$
where
$
y'_{M_{12}}=\frac{\sqrt{-M_{12}^2}}{\sqrt{4m^2-M_{12}^2}}.
$

Finally, the equation for the Faddeev amplitude reads:
\begin{equation}\label{eq29a}
\Gamma_{12}(R_{\perp},x)
=F(M_{12})\frac{\displaystyle{1}}{\displaystyle{(2\pi)^3}}
\displaystyle{\int_0^1}
\displaystyle{dx'}
\displaystyle{\int_0^{\infty}}
\frac{\Gamma_{12}\left(R'_{\perp},x'(1-x)\right)\;d^2R'_{\perp}}
{\displaystyle{(\vec{R'}_{\perp}-x'\vec{R}_{\perp})^2+m^2-x'(1-x')M_{12}^2}}.
\end{equation}
The three-body mass $M_3$ enters in this equation through variable
$M_{12}^2$, defined by (\ref{eq17}).

By replacing  $x'(1-x)\to x'$, equation (\ref{eq29a}) can be transformed to 
\begin{equation}\label{eq30}
\Gamma_{12}(R_{\perp},x)
=F(M_{12})\frac{\displaystyle{1}}{\displaystyle{(2\pi)^3}}
\displaystyle{\int_0^{1-x}}
\frac{\displaystyle{dx'}}{\displaystyle{x'(1-x-x')}}\;
\displaystyle{\int_0^{\infty}}
\frac{\displaystyle{d^2R'_{\perp}}}{\displaystyle{{{\cal M}'}^2-M_3^2}}\;
\Gamma_{12}\left(R'_{\perp},x'\right),
\end{equation}
where
\[ {{\cal M}'}^2=\frac{\vec{R'}^2_{\perp}+m^2}{x'}
+\frac{\vec{R}^2_{\perp}+m^2}{x}
+\frac{(\vec{R'}_{\perp}+\vec{R}_{\perp})^2+m^2}{1-x-x'}. \]

This equation is the same than equation (11) from \cite{tobias} except
for the integration limits of ($\vec{R'}_{\perp},x'$) variables.
In \cite{tobias} the integration limits follow from the condition $M_{12}^2>0$.
They read
\begin{equation}\label{FBC}
\int_{m^2\over M_3^2}^{1-x} \left[\ldots\right] dx'
\int_0^{k^{max}_{\perp}} \left[\ldots\right]  d^2R'_{\perp}
\end{equation}
with $k^{max}_{\perp}=\sqrt{(1-x')(M_3^2x'-m^2)}$
and introduce a lower bound on the three-body mass $M_3>\sqrt{2}m$.
The same condition,
though in a different relativistic approach, was used in \cite{noyes}.
The integration limits in (\ref{FBC}) restrict the arguments
of $\Gamma_{12}$ to the domain
\[ {m^2\over M_3^2} \leq x \leq 1 -{m^2\over M_3^2} ,
\quad 0\leq R_{\perp}\leq k_{\perp}^{max}\]
and can be considered as a  method of regularization.
In this case, one no longer deals with the zero-range forces.

Being interested in studying the zero-range interaction,
we do not cut off the variation domain of variables $R_{\perp},x$
\[0\leq x\leq 1,\quad 0\leq R_{\perp}< \infty.\]
The integration limits for these variables 
reflect the conservation law of the
four-momenta  in the three-body system and they are automatically fullfilled, as
far as the $\delta^{(4)}$-function in  (\ref{eq3}) is taken into account.
The off-shell variable $M_{12}^2$ may take negative values,
when $R_{\perp}$ and $x$ vary in their proper limits.
Thus, if $M_3^2>m^2$ one has $-\infty \le M_{12}^2 \le (M_3-m)^2$
but if  $M_3^2<m^2$,  $M_{12}^2$ is always negative $-\infty \le M_{12}^2 \le 0$.
We would like to notice that $M_{12}^2$
is not to be confused with the on-shell effective mass squared
$s'_{12}=(k'_1+k'_2)^2$
which is indeed always positive and even $s'_{12}\geq 4m^2$.
As we will see, this point turns out to be crucial for the 
appearence of the relativistic collapse.

\begin{figure}[h]
\begin{center}
\epsfxsize=8.8cm\subfigure[ ]{\epsffile{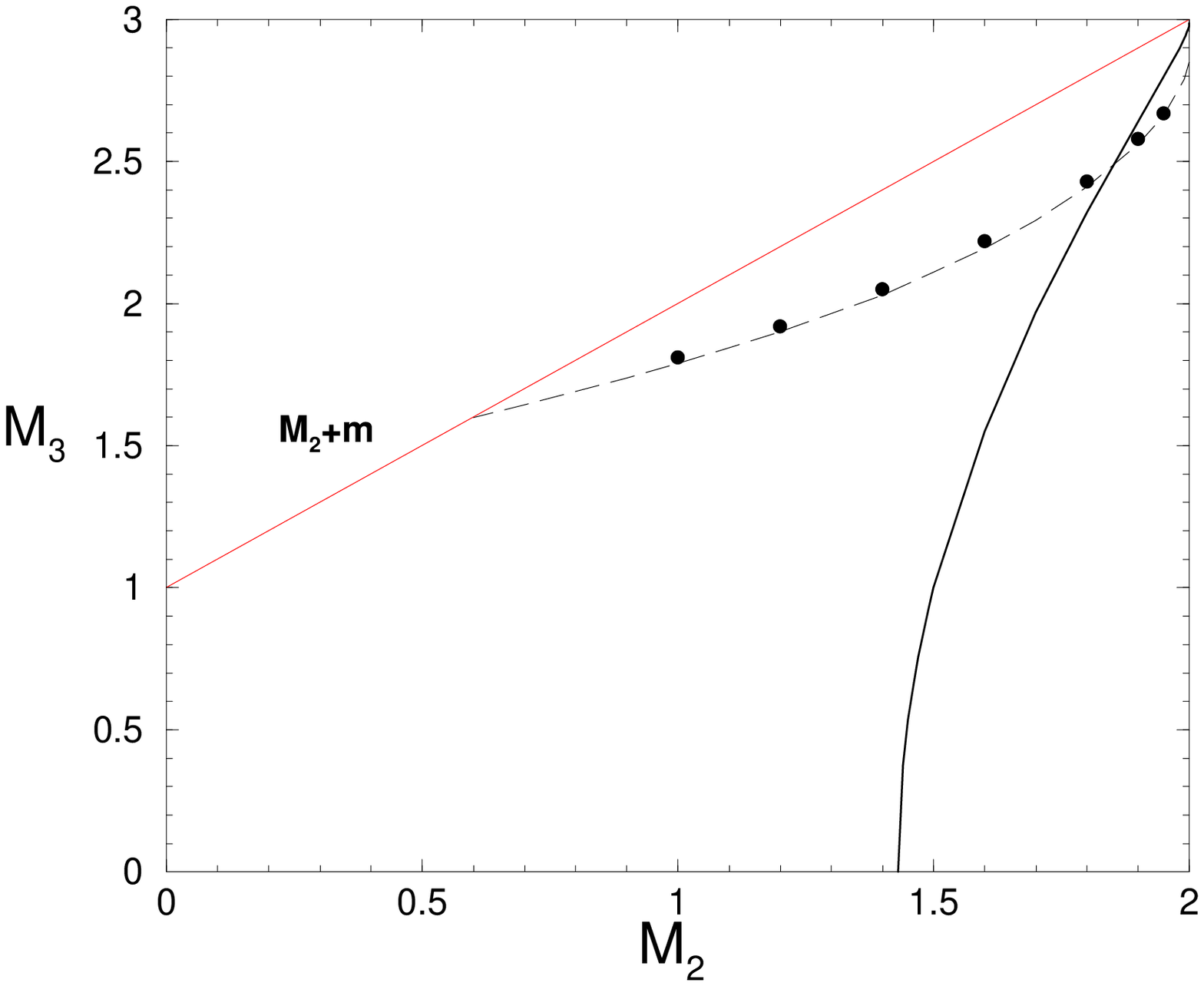}}\hspace{0.5cm}
\epsfxsize=6.8cm\subfigure[ ]{\epsffile{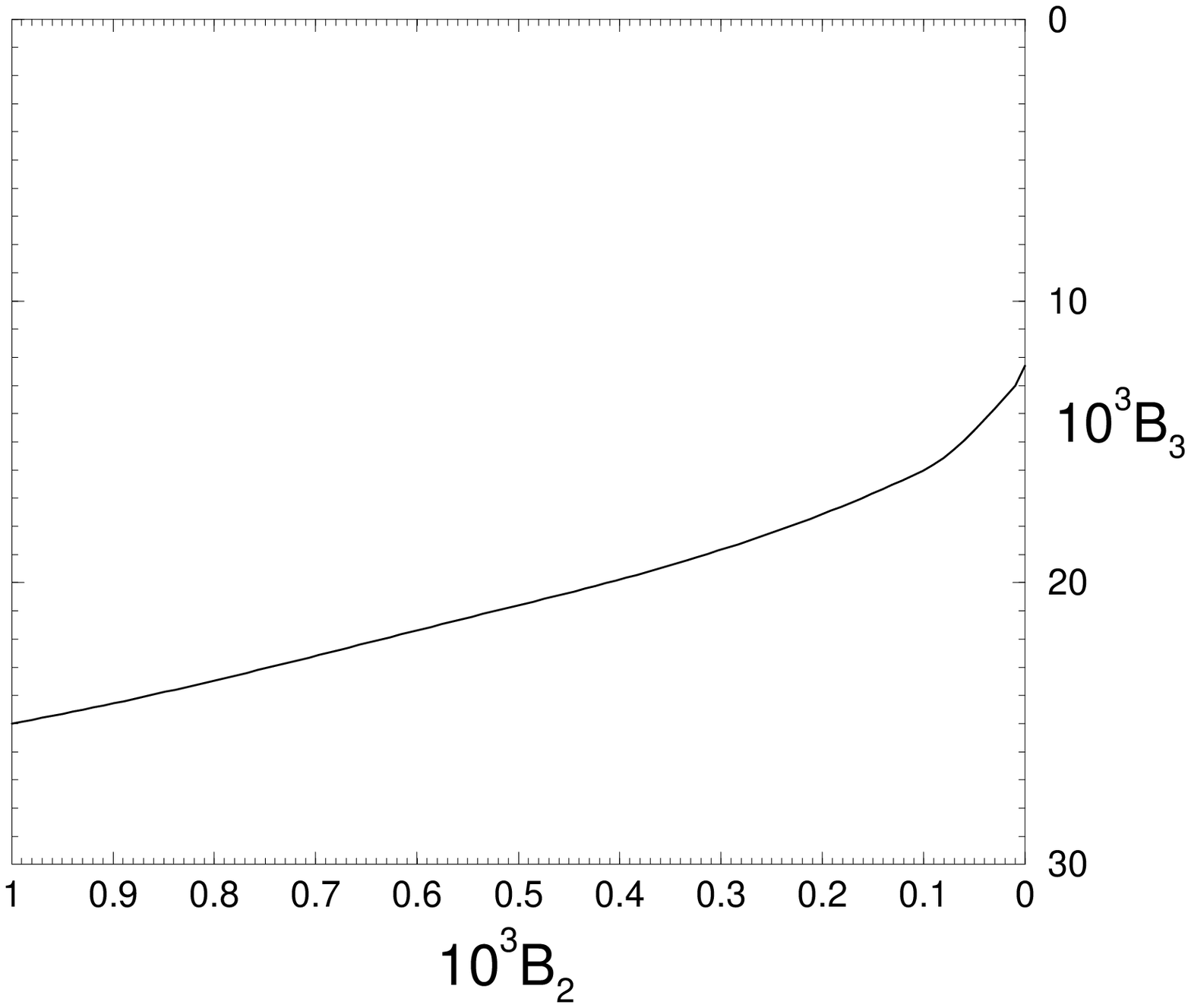}}
\caption{(a) Three-body bound state mass $M_3$ versus
the two-body one $M_{2}$ (solid line). 
Results  obtained with
integration limits (\protect{\ref{FBC}}) are in dash line.
Dots values are taken from \cite{AMF_PRC_95}.
(b) Zoom of the zero two-body binding region ($M_2\to2m,B_2\to0$)
displaying solid line only.}\label{Fig_M3_M2}
\end{center}
\end{figure}

The results of solving equation (\ref{eq29a}) are
presented in what follows. Calculations were carried out with constituent mass 
$m=1$ and correspond to the ground state.
We represent in fig. \ref{Fig_M3_M2}a
the three-body bound state mass $M_3$ as a function of the two-body one $M_2$
(solid line) together with the dissociation limit $M_3=M_2+m$.
The zero two-body binding limit $B_2=2m-M_2\to0$
is magnified in fig. \ref{Fig_M3_M2}b. In this limit
the three-boson system has a binding energy  $B_3\approx0.012$.

When $M_2$ decreases, the three-body mass $M_3$  decreases very
quickly and vanishes at the two-body mass value $M_{2}=M_{c}\approx 1.43$.
Whereas the meaning of collapse as used in the Thomas paper implies unbounded
nonrelativistic binding energies and cannot be used here,
the zero bound state mass $M_3=0$ constitutes its relativistic counterpart.
Indeed, for two-body
masses below the critical value $M_c$, the three-body system  no longer exists.
\begin{figure}[!ht]
\begin{center}
\mbox{\epsfxsize=7.cm\epsffile{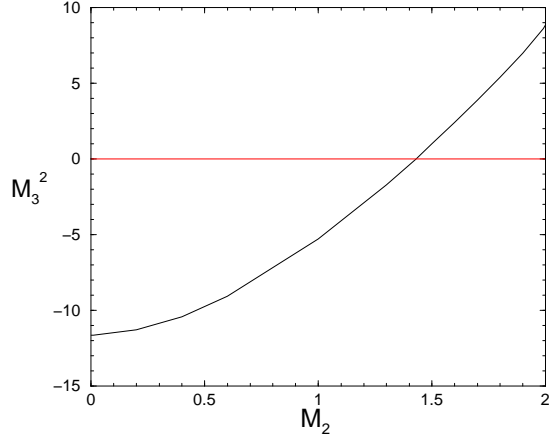}}
\caption{Three-body bound state mass squared $M_3^2$ versus $M_{2}$.}
\label{Fig_M32_M2}
\end{center}
\end{figure}

The results corresponding to integration limits (\ref{FBC})
are included in fig. \ref{Fig_M3_M2}a (dash line) for comparison.
Values given in \cite{tobias} were not fully converged \cite{PC}.
They have been corrected in \cite{AMF_PRC_95} and are indicated by dots.
In both cases the repulsive relativistic effects produce a natural cutoff in
equation (\ref{eq29a}), leading to a finite spectrum and -- in the Thomas sense
-- an absence of collapse, like it was already found in \cite{noyes}.
However, except in the zero binding limit, solid and dash curves strongly differ from each other.

We would like to remark that for $M_2\leq M_c$, equation (\ref{eq29a}) posses
square integrable solutions with negative values of $M_3^2$. They have no
physical meaning but $M_3^2$ remains finite in all the two-body mass range
$M_2\in[0,2]$. The results of $M_3^2$ are given in figure \ref{Fig_M32_M2}.
When $M_{2}\to 0$,  $M_3^2$ tends to $\approx -11.6$.

\begin{figure}[h]
\begin{center}
\vspace{-.8cm}
\epsfxsize=7.5cm\subfigure[ ]{\epsffile{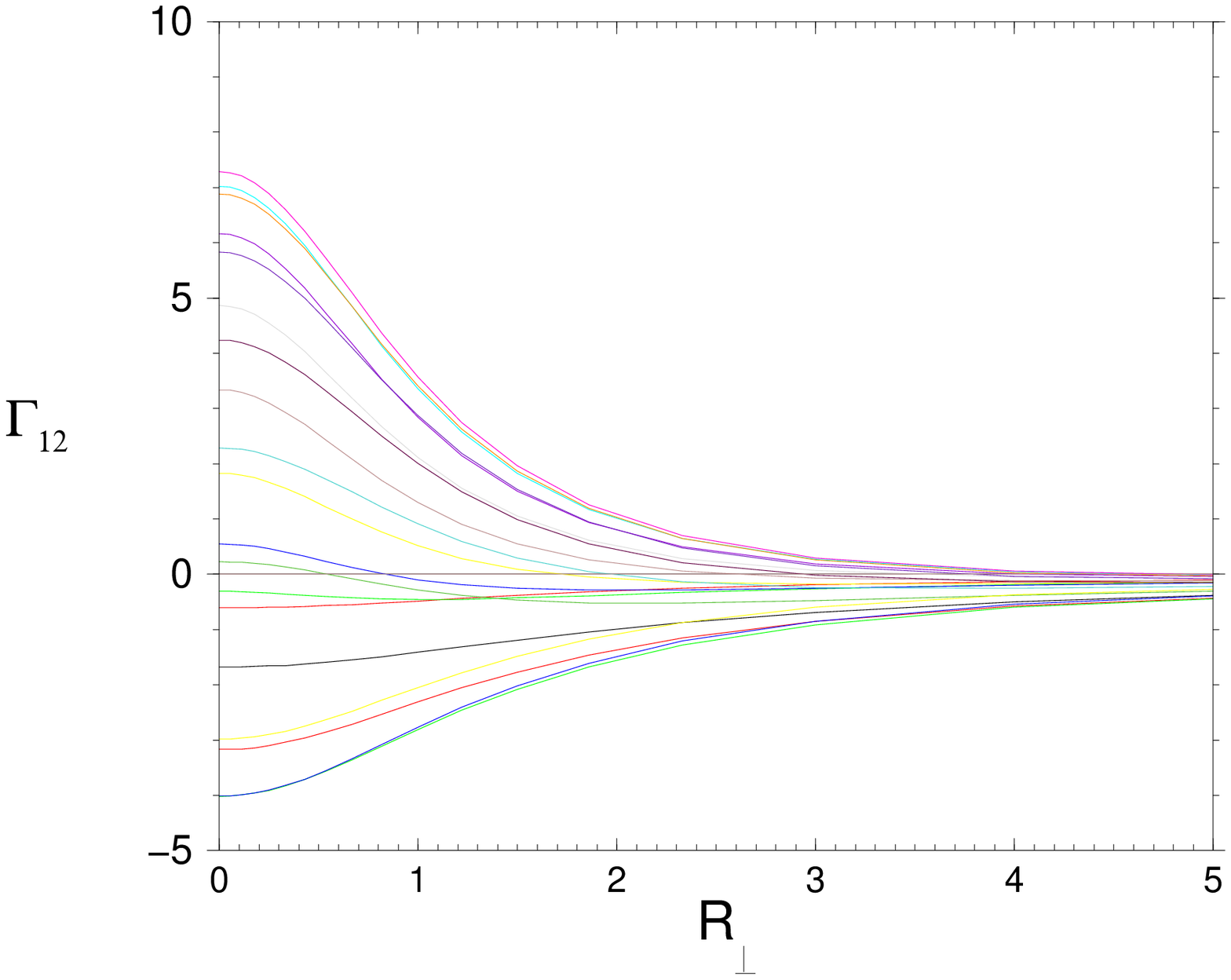}}\hspace{0.5cm}
\epsfxsize=7.5cm\subfigure[ ]{\epsffile{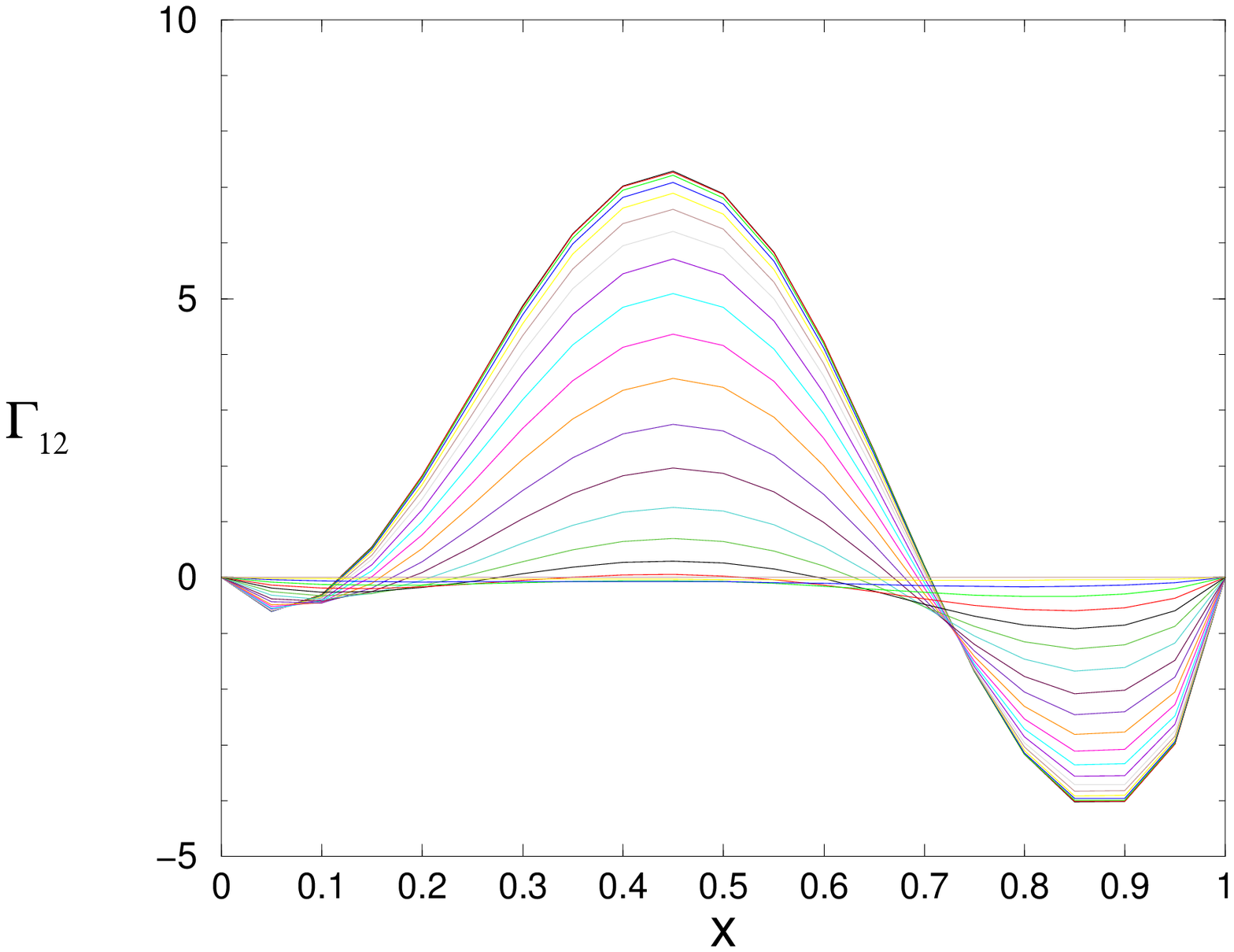}}
\vspace{-.cm}
\caption{Faddeev amplitude $\Gamma_{12}$ for $M_2=1.5$, $M_3=1$ 
(a) as a function of $R_{\perp}$ at fixed values of $x$  and 
(b) as a function of $x$  at fixed $R_{\perp}$ values.}\label{G_1.5_1}
\end{center}
\end{figure}

In figure \ref{G_1.5_1} is shown the Faddeev amplitude $\Gamma_{12}$
corresponding to a solution with relatively large binding energies
$M_2=1.5$ and $M_3=1$. Figure \ref{G_1.5_1}a shows its
$R_\perp$-dependence at fixed values of $x$ and figure \ref{G_1.5_1}b --
its $x$-dependence at fixed values of $R_\perp\in[0,20]$.
One can remark in figure \ref{G_1.5_1}b that, in addition to the zeroes at
$x=0$ and $x=1$, $\Gamma_{12}(R_\perp,x)$ has two zeroes at 
$x\approx 0.15$ and $x\approx0.7$.  
When the mass values $(M_2,M_3)$ are
changed, the positions of zeroes are only slightly shifted.

In summary, we have considered the relativistic problem of three equal-mass bosons,
interacting via zero-range forces constrained
to provide finite two-body mass $M_{2}$. The
Light-Front Dynamics equation has been derived and solved numerically.

We have found that the three-body bound state exists 
for two-body mass values in the range $M_c\approx 1.43\,m \leq M_{2}\leq 2\,m$.
At the zero two-body binding limit, the three-body binding energy is
$B_3\approx 0.012\,m$.
The Thomas collapse is avoided in the sense that three-body mass $M_3$ 
is finite, in agreement with \cite{noyes,tobias}.
However, another kind of catastrophe happens.  
Removing infinite binding
energies, the relativistic dynamics generates zero three-body mass
$M_3$ at a critical value  $M_2=M_c$. 
For stronger interaction, i.e. when $0\leq M_{2}< M_c$, 
there are no physical solutions with
real value of $M_3$.
In this domain, $M_3^2$ becomes negative and the three-boson system
cannot be described by zero range forces, as it happens
in nonrelativistic dynamics. 
This fact can be interpreted as a relativistic collapse.

\bigskip
{\bf Acknowledgements.}
Authors are indebted to T. Frederico for useful remarks
concerning results of Refs. \cite{tobias,AMF_PRC_95}.
V.A.K. is sincerely grateful for the warm hospitality of
the theory group at the Institut des Sciences Nucl\'{e}aires
in Grenoble, where this work was performed. 
This work is
partially supported by the French-Russian PICS and RFBR grants Nos. 1172 and
01-02-22002.
Numerical calculations were performed  
at CGCV (CEA Grenoble) and  IDRIS (CNRS).


\end{document}